# Single cell visualization of transcription kinetics variance of highly mobile identical genes using 3D nanoimaging

Paolo Annibale

pannibal@uci.edu

Enrico Gratton*

*egratton22@gmail.com

Laboratory for Fluorescence Dynamics, Department of Biomedical Engineering,

University of California Irvine

**Abstract**

Multi-cell biochemical assays and single cell fluorescence measurements revealed that the elongation rate of Polymerase II (PolII) in eukaryotes varies largely across different cell types and genes. However, there is not yet a consensus whether intrinsic factors such as the position, local mobility or the engagement by an active molecular mechanism of a genetic locus could be the determinants of the observed heterogeneity. Here by employing high-speed 3D fluorescence nanoimaging techniques we resolve and track at the single cell level multiple, distinct regions of mRNA synthesis within the model system of a large transgene array. We demonstrate that these regions are active transcription sites that release mRNA molecules in the nucleoplasm. Using fluctuation spectroscopy and the phasor analysis approach we were able to extract the local PolII elongation rate at each site as a function of time. We measured a four-fold variation in the average elongation between identical copies of the same gene measured simultaneously within the same cell, demonstrating a correlation between local transcription kinetics and the movement of the transcription site. Together these observations demonstrate that local factors, such as chromatin local mobility and the microenvironment of the transcription site, are an important source of transcription kinetics variability.

**Introduction:**

A growing evidence points to a significant variance characterizing the transcription process[1]. Sizable cell-to-cell differences were recently quantified in the amount of transcript of identical genes[2] and stochastic gene expression from an isogenic cell line containing a single fluorescent reporter gene was measured to depend upon the genomic site of the insertion[3]. Although elongation is one of the fundamental steps of transcription, and it is now accepted that elongation rate may play a regulatory function, the exact determination of the speed of the elongation complex processivity in vivo is still the subject of debate[1]. Recent global run-on sequencing experiments demonstrated that elongation rates varies up to four times among distinct genomic loci, and that PolII can elongate at different speeds along identical genes in different cell lines and in response to different signaling pathways[4]. This was recently confirmed by a fluorescence microscopy study that highlighted a significant variability in PolII elongation rates along the MDN1 gene, although measured across different yeast cells[5].

However, conclusive evidence able to identify whether factors such as genomic position, local mobility and microenvironment, are indeed the primary determinant of the observed variability in the kinetics of transcription, in particular elongation, is missing. In particular, the question whether increased local mobility of individual chromatin regions correlates to their transcriptional output is still the subject of active research[6]. Studies which exploited heat-sensitive PolII mutants in yeast did not observe any change in the confinement radius of fluorescently tagged loci in cells were elongation was disabled[7]. In mammalian systems the transcriptional activity of selected genes was recently correlated to their physical displacement within the nucleus, but only in the particular case of transcriptional activation following an actin dependent, long range motion (0.5-6 µm) of HSP70 genes towards the nuclear speckles after heat shock[8].

The reasons for the large variability affecting PolII transcription have been so far difficult to isolate, even in single cell experiments. To do so, it would be necessary to first isolate those extrinsic factors, such as cell to cell variability, that contribute a large part of this variance. This would allow assessing the role of the dynamic, spatial properties of the active site in regulating transcription kinetics. We propose here an original application of a fast 3D nanoimaging method, namely 3D orbital particle tracking, towards this goal. We sought to exploit the observation of a spatial structure in the organization of fluorescently labeled nascent mRNAs within the model system represented by the transgene array of U2OS 263 cells, a cell line allowing simultaneous labeling of both a gene and its transcript[9]. We observe that mobile mRNA foci or *petals* surround the denser chromatin array.

Although transgene arrays are artificial systems, they have been highly valuable in the past to study the dynamics of large scale chromatin domains[10, 11, 12, 13] and its overall transcription[14]. Here we examine specifically the transcription dynamics of individual genes within the repeat and its relationship with chromatin mobility in vivo. Each copy of the 200 gene repeats composing of the transgene array can be visualized by means of a Lac Operator cassette[15] and an MS2 cassette allows nascent mRNA to be decorated by a fusion of EGFP with MS2, a coat protein which recognizes a specific and repeated hairpin sequence in the mRNA molecule[16]. Upon Doxycycline (+Dox) induction it was demonstrated that the array transitions from a heterochromatic state to a euchromatic condition following histone exchange that removes H3K9 methylation[9]. The authors of the study provided evidence that only parts of the repeats were being transcribed, suggesting higher resolution analysis of the locus to determine effects of sub chromosomal features on kinetics of mRNA synthesis.

We developed a way to visualize the spatial and fluorescence trajectory of each of the petals at high resolution over space and time by means of a fluorescence nanoimaging method based on 3D orbital tracking in a 2-photon laser-scanning microscope. 3D orbital tracking is ideally suited to capture fluctuations in the fluorescence signal arising from the enzymatic activity of PolII within the reference system of the active fluorescently labeled array at ms temporal resolution. In combination with phasor analysis, a fit-free method that allows us visualizing and comparing PolII elongation rates within different petals, we measure a fourfold variation in the average elongation rates across identical copies of the same gene measured simultaneously within the same cell.

Furthermore, the position of individual fluorescent particles within the orbit can be determined with nm-accuracy, depending exclusively on the measured signal to noise ratio [17] [18]. This allows us determining if any relationship exist between the kinetic rates measured in the petals and their local mobility. We observe that those petals that can explore a larger region of space also display higher elongation rates. Finally, by cross-correlating the nm-sized displacement of petals to the concomitant biogenesis of mRNA molecules, we provide evidence of an active mechanism determining directional motions of the active loci concomitant to increases in transcription.

**Results:**

**Active transcription sites form a dynamic petal-like structure upon induction**

We found that the mRNA fluorescence signal, following induction in U2OS 263 cells, concentrates in multiple well-defined regions surrounding the denser core of the labeled transgene array (**Figure 1a)**, of approximately one μm in diameter**.** The number of these complexes varies from cell to cell, from a lower number of one, up to seven distinguishable loci. After the first ten to fifteen minutes following induction, the fluorescence from the petals reaches a steady state that is maintained over a timescale of hours. Each petal undergoes sizable angular displacements in the reference system centered on the gene array. 2D time-lapse imaging **(Supplementary Movie 1)** clearly shows that the radial position of this mRNA corona is very stable, while the petals tend to move on the surface of the chromatin array. The fluorescently labeled gene array also displays a highly dynamic behavior, owing to overall chromatin motion, displacement of the nucleus, global movement of the entire cell and μm-sized sample drifts occurring during the experiments **(Supplementary Figure 1)**.

In order to capture the rapid dynamic processes occurring in each petal and to decouple them from the overall motion of the gene array we employed 3D orbital tracking to follow the labeled transgene array at millisecond temporal resolution and with nm precision in a 2-photon laser scanning microscope **(**see **Materials and Methods** and description of the setup**)**[19]. Briefly, as illustrated in **Figure 1b,** the tightly focused beam of a pulsed infrared femtosecond laser light source is steered continuously in circular orbits around the fluorescently labeled transgene array. As opposed to raster scan imaging, this approach improves significantly the temporal resolution of the measurement and each orbit can be centered on the recovered object position using an on-the-fly localization and feedback mechanism[17]. Detection in multiple colors allows using one spectral channel for tracking the transgene array and the other to follow the fluorescence trajectory of the MS2 decorated mRNA petals. Since the PSF of the microscope is elongated along the axial direction, and furthermore the orbits are performed at a certain distance ($\sim 1$ μm) from the center of the transgene array, the contribution of axial fluctuations of the petals or of the residual radial motion are almost negligible **(Supplementary Figure 2)**. The fluorescence collected along multiple orbits can then be displayed in the form of a kymograph or intensity carpet, and in this representation each petal appears as a vertical trace, as displayed in **Figure 1c**. Intermittency in the fluorescence on both fast and slow timescales is visible.

Each petal can be individually resolved by the microscope PSF, and its angular trajectory over time tracked continuously by fitting the carpet with a Gaussian function line by line[20]. This process allows constructing kymographs corrected for any spatial fluctuation, and the remaining fluorescence fluctuations arise only because of kinetic processes, such as increases in fluorescence signal due to the addition of MS2 fluorescent subunits as PolII elongates, or decreases in fluorescence due to the release of fluorescently labeled mRNA molecules **Figure 1c-d**. Furthermore, the values used to correct the position of the petals produce trajectories of the center of the petal with nm-precision, depending only upon the signal to noise ratio. It is then possible to reconstruct both the fluorescence and spatial trajectory of each of the petals over an unprecedented period of time (up to hours) and temporal resolution (260 ms in this case), as displayed respectively in **Figure 1e-f.**

**The observed steady state is a dynamic equilibrium determined by the biogenesis of new mRNA molecules and the release of complete mRNAs in the nucleoplasm**

We found that fluorescence trajectories are not spatially or temporally cross correlated (data not shown) among petals surrounding the same chromatin locus, supporting the notion that each petal on the surface of the denser chromatin array acts as an independent locus. These observations argue in favor of the possibility of measuring the transcriptional activity of distinct petals within the transgene as an interesting model of transcription from tandemly repeated genes[21]. Upon photo bleaching, as illustrated in **Supplementary Figure 3**, it is possible to observe a complete recovery of the fluorescence intensity of a petal, confirming that these structures are regions of active transcription and not of local accumulation of mRNA transcript. To achieve equilibrium, the synthesis of new molecules needs to be compensated, to avoid ongoing accumulation of mRNAs on the petal, which we do not observe. This suggests that fluorescent mRNA molecules are not only synthesized, but also released from the petals in the nucleoplasm, resulting in a dynamic equilibrium at the locus.

We validated this hypothesis by performing tracking-Pair Correlation Analysis, a fluorescence fluctuations microscopy method that provides information on the degree of molecular diffusion or flow between two points along the path of a fast scanning laser beam[22]. The pair correlation method is based on spatial cross-correlation in the fluorescence intensity arising from two distinct points at a distance *d*. A cross correlation signal indicates that molecules that leave one point can reach the other. The delay in such spatial cross-correlation reflects the average physical length of the path taken by a molecule. Starting from the fluorescence intensity carpet collected along the trajectory of the scanning beam, a new carpet can be therefore constructed: the intensity values in each column is indicative of diffusion or flow from that point to a point at a distance $d$ along the orbit in the direction followed by the scanning beam[22][23].

As illustrated in Figure 2**a**, for this measurement we switched from a circular to a trefoil orbit that allows us measuring diffusion and flow of mRNA molecules across different paths surrounding the transgene array[24]. We focused in particular on the motion from the petals far into the nucleoplasm (point 1 to 2, 4 to 6 and 10 to 11 in the Figure 2**a** diagram), which robes the the ability of mRNA to leave or re-enter the core of the transgene array (Figure 2**a**  3 to 5 and 8 to 9 respectively). Each of these pairs of locations is spaced 12 pixels along the orbit. Figure 2**b** displays the pCF carpet calculated using a $d$=12 pixels distance between the columns of the intensity carpet. A strong spatial cross-correlation is observed between the petals and the distal extremities of the orbit lobes, confirming that most of the mRNA molecules that reach further into the nucleoplasm originate from the transcribing petals. Barriers to diffusion, in the form of dark pCF carpet columns, are instead observed for displacements towards or away from the center of the array, confirming that mRNA synthesis occurs on the surface and that the chromatin density within the array impedes any mRNA diffusion or flow. Finally, mRNA nucleoplasmic diffusion is detected at the distal extremities of the trefoil lobe. Interestingly, a significant difference can be observed in the molecular flow of mRNA molecules that originate from each of the three petals. Figure 2**c** illustrates that the average delay experienced by mRNA molecules leaving the three positions marked as 1,4 or 10 in reaching the nucleoplasm is different[25].

**Extracting PolII transcription kinetics using fast fluorescence fluctuations and the phasor approach**

We use fluctuation correlation spectroscopy in combination with the phasor analysis to extract transcription kinetics, and in particular PolII elongation kinetics within the petals. As illustrated in **Supplementary Figure 4a-b**, the fluorescence trajectory of a petal can be collected for up to hours following induction, displaying fast fluctuations in the 0.1-10 s timescale as well as much slower intensity fluctuations over a timescale ranging from tens of seconds to several minutes.

This observation is in qualitative agreement with recent observations of the dynamics of fluorescently labeled mRNA on a single transcription site[26] [27]. Furthermore, these fast fluctuations are temporally correlated, as observed calculating the AutoCorrelation Function (ACF) of the time series (**Supplementary Figure 4c**). The timescale of the correlation reflects the underlying kinetics of the transcription process (**Supplementary Figure 5**).

To extract elongation kinetics we adopt here a novel strategy, based on the calculation of the phasor of the ACF. Phasor analysis, i.e. the representation of the harmonic components of a signal as a point on the unit circle in the complex plane, was successfully applied to study lifetime data[28, 29], spectral images[30, 31], and more recently diffusion[32]. It allows for a prompt visualization and fingerprinting of the data from a time series, providing an intuitive and graphical way to distinguish multi-exponential lifetime decays, closely spaced spectra and, when applied to the ACF of a short fluorescence time series and referenced using simulated data, molecular diffusion coefficients. Since the experimental ACF displays an exponential character (as illustrated in **Supplementary Figure 3c**), the analogy to the phasor treatment of exponentially decaying fluorescence lifetimes is immediate.

Using this method, as is schematically illustrated in **Supplementary Figure 6**, each petal kymograph maps to a cloud of points upon the unit circle, and its average angular coordinate reflects the kinetic information contained within the fluorescence fluctuations. Angular values are converted into elongation rates using a reference set of simulated PolII elongation runs with different velocities as well as different termination rates. A full description of the simulation code and the phasor analysis can be found in the **Supplementary Materials and Discussion**.

The parameters for the simulations were taken from Darzacq *et al* [14]. If, for example, we assume an effective elongation rate of approximately one MS2 loop/s (corresponding to about 60-65 bp/s), and a mRNA release time of about 500 s due to 3' end processing, the dwell time of fluorescently labeled mRNA on the transcript would be of the order of 550 s. Ideally, for an individual transcript, intensity fluctuations should not be observed any more during the release time, i.e. after PolII elongation on the 24x MS2 cassette (**Supplementary Figure 5**). Therefore, an analysis that high-pass filters the signal for fluctuations occurring on a faster timescale, such as dividing the time trace in bins, and calculating the ACF and the phasor for each bin, would be able to highlight fluctuations arising from the elongation step.

We validated this assumption by generating simulated trajectories representing the fluorescence time traces of a pool of PolIIs (*N=100*) initiating, elongating and being released from an individual gene (see **Supplementary Materials and Discussion**) at elongation rates ranging from a few bp/s and including previously reported PolII elongation values at 12 bp/s[33], 20 bp/s[26], 60 bp/s[14], 120 bp/s, and finally up to a much higher rate of 600 bp/s[34]. Termination times were also varied over two orders of magnitude, from about 5 s to 500 s. Phasor analysis of the simulated datasets highlights that individual changes in transcription elongation give rise to a significant change in the corresponding phasor (**Supplementary Figure 7a-b**), and that, as expected, elongation changes can be detected unambiguously over at least an order of magnitude (8-80 bp/s) even in presence of the large variance in termination rates. Changes in initiation rate did not appear to influence the analysis once a steady state was reached within the petal (data not shown). On the other hand, if the termination rate is assumed comparable to the elongation rate (as done in Larson *et al.*[26]), the resulting phasor-angle vs termination allows measuring elongation changes over two orders of magnitude (**Supplementary Figure 7c**). Phasors of the experimental trajectories can then be plotted on the unit circle and referenced using the simulated ones, as illustrated in (**Supplementary Figure 7d**).

**Phasor analysis highlights a wide range of transcription kinetics within a single cell**

The phasor approach allows prompt comparison of elongation velocities across different cells, across different petals within the same cell, and even between temporally distinct portions of the same petal. Figure 3 displays the PolII elongation rates extracted from the phasor analysis performed on all the petals kymographs of each of 11 interphase U2OS 263 cells following Dox induction. Each cell (labeled *A-K)* displays a quite distinct average elongation value, pointing to the fact that PolII elongation on the same gene displays a large variability across different cells[5]. As illustrated by the scatter cloud indicated by *ALL* in Figure 3**a**, the combined PolII elongation values span almost entirely the range of elongation speeds reported in the MS2 literature, from below 10 bp/s to over 200 bp/s, according to our calibration[1]. The average elongation value that we measure (42 bp/s) is in excellent agreement with the recent measurement from run-on sequencing, which yields 35 bp/s[4].

Since an individual array gives rise to multiple petals, the tracking-phasor analysis allows comparing the transcription kinetics of multiple, identical genetic loci within the same cell and at the same time. Figure 3**b** illustrates that within an individual cell (e.g. cell *E*) PolII elongation rate can change of more than a factor of four between the slowest and fastest petal, while the variance of elongation rates measured within a petal can span over an order of magnitude.

Furthermore, our approach allows measuring the temporal evolution of PolII elongation within an individual petal over time. Since phasor analysis divides the carpet in periods, corresponding to a distinct time, it is possible to observe the evolution of PolII elongation rate along a specific petal. Figure 3**c** displays the example of a petal "switching on" following Dox induction. The markers in Figure 3**d** display the corresponding PolII elongation rate and are color coded according to time. This approach reveals that, as the petal turns on, the average PolII elongation rate increases by almost an order of magnitude over an interval of 45 minutes. Figure 3**e** displays the reverse situation, where the transcriptional activity of a steadily expressing petal is inhibited using the drug ActinomycinD (AD). Following perfusion with the drug, the measured PolII velocity drops, again by almost an order of magnitude, as indicated in Figure 3**f.** Together these results indicate that PolII elongation along a gene can significantly change over time, and that the variance observed for each of the petals is compatible with a range of PolII velocity between extremes as far apart as an order of magnitude.

**Active transcription correlates with physical movements of the transcription sites**

The capability to perform observations on a single loci as a function of time with high spatial and temporal resolution prompted us to address the longstanding question whether a relationship exists between transcriptional activity and the physical movements of the inducted gene [6]. Our nanoimaging method provides both the spatial and temporal resolution necessary to address this question providing novel information: nm-sized displacements of the active genes in the reference system of the array, and therefore devoid of any drift, can be correlated at sub-second temporal resolution to the fluorescence fluctuations arising from mRNA MS2-EGFP binding to newly synthesized mRNA molecules as PolII elongates.

In the polar coordinates reference system centered on the transgene array, the motion of the petals is predominantly angular, as can be observed in Figure 4**a** which displays the representative positions of petals observed over time by 2D time lapse imaging in a plane crossing the transgene array. Each petal typically explores an arc of a few hundred nanometers along the circumference of the array (the azimuthal angle $\varphi$. The polar angle $\theta$ is also expected to change, but its changes are not detected in the kymographs nor they influence the kymograph intensity given the multiplane orbital tracking configuration and the geometry of the PSF, as illustrated in Supplementary Figure 2)**.** We then calculated the MSD of the angular ($\varphi$) trajectories of each petal, in units of $\mu$m along the circumference of the transgene. The MSD analysis has been routinely employed in the analysis of trajectories of individual chromosomal loci: the constraint radius $R_{max}$ measured from the saturation value of the MSD is a common indicator of chromatin mobility [35] and the functional dependence of the MSD curve at different timescales was employed to investigate chromatin properties in the context of polymer-like fibers [36].

Figure 4**b** displays the experimental MSD curves calculated using a lag time $\tau$ covering an interval between 1 s and 512 s. The average MSD curve displays clearly the presence of two regimes, exhibiting a power law dependence on time: a short time regime (<5 s) with an exponent equal to 0.99 ± 0.04 and an intermediate regime (10-100 s) with an exponent equal to 0.38 ± 0.01.

Power Law fits, in the form $y = At^\alpha$ were performed for each MSD curve in each of the two regimes (0-5s) and (10-100s). In the first regime the exponent $\alpha$ was found to vary between a minimum of 0.31 and a maximum of 1.43 and in the second regime between a minimum of 0.09 and a maximum of 0.46.

Simulations of chromatin dynamics[36] predict that the MSD of a diffusing chromatin fiber crosses over from a power law behavior with $\alpha$=0.75 to one with 0.25<$\alpha$<0.5. This is graphically illustrated by the dashed guides for the eye in Figure 4b, which display the slope of these limiting power law behaviors. Our results are in qualitative agreement with these simulated values, and they provide evidence that the observed motion of the petals is compatible with the diffusion of chromatin fibers.

The MSD analysis allows establishing a direct correlation between the overall mobility of individual petals and the elongation rate measured using the phasor analysis.

For most of the petals no obvious saturation of the MSD curve was observed at longer times, the most likely explanation being slow rotations of the transgene array itself on a timescale longer than a few hundred seconds. We therefore arbitrarily define $R_{max}$ as the extrapolation of the power law behavior observed in the 10-100 s regime at t=512 s. Although strictly speaking this is not a radius of constraint, since no constrained motion is ultimately observed for most petals, $R_{max}$ nonetheless provides an indication of the spatial range that can be explored by the petals on a timescale of approximately 10 minutes. The average radius of constraint $R_{max} = (5/2 \cdot MSD)^{0.5}$ extrapolated from the intermediate regime to $\tau = 512$ s) is about 210 nm, and the petals that display a lower overall mobility (small $R_{max}$ values) appear to be elongating at a lower rate than the more mobile ones, as illustrated by the scatter plot reported in Figure 4**c** (Pearson *R* =0.52, *p-value*=0.008).

Further light can be shed on whether an active mechanism is indeed responsible for the motion of the transcribing petals. By cross-correlating the fluorescence intensity of each petal at each time to its angular displacement we found that changes in transcript output correlate to movements of the active transcription site. The diagrams in Figure 4**d-e** illustrate that the sign of the cross-correlation function in Figure 4**e** depends upon the direction of motion of the petal. The cross-correlation carpet (Figure 4**f,** normalized to 99% confidence bands calculated for pure diffusion of the petals) highlights the transition from negative to positive cross-correlation values, after a delay typically ranging from tens to hundreds of seconds. This observation is a strong indication that (i) there is a prevalent direction in the angular fluctuations of the petals (either clockwise or counterclockwise in the reference system of the laser orbit), and (ii) increasing fluorescence in the petal correlates with a directional movement for up to about 10 s, to be followed by a recoil of the petal in the other direction. It should be emphasized here that the cross-correlation analysis reported in Figure 4**f** highlights different information than the MSD analysis, and while the overall degree of chromatin movement extracted from the $R_{max}$ values analysis provides an indication for limiting long time scales, the cross-correlation indicates that directional movements occur with a specific delay with respect to maxima of transcriptional activity. This result can be reconciled only with the existence of a molecular mechanism driving the displacement of the active transcription site, possibly in a spring and ratchet mechanism [37].

This is consistent with the drastic reduction observed in the amplitude of the angular fluctuations upon energy starvation of the cells achieved using Sodium Azide and 2-DeoxyGlucose as illustrated in **Supplementary Figure 9**, although a local reduction of overall chromatin mobility could be a competing factor for this observation. ATP depletion was demonstrated to cause a change in nuclear substructure leading to a rearrangement of nuclear dense structures that affects intranuclear diffusion of mRNAs and may ultimately limit the diffusive behavior of chromatin itself.

**Discussion**

Here we have shown that we can follow continuously for up to hours and at an unprecedented ms temporal resolution both the transcriptional activity and spatial trajectory of identical active genes, simultaneously within the same cell. This is made possible by the peculiar, and previously underappreciated arrangement of active transcription sites in U2OS 263 cells: a corolla of active transcription regions surrounding the compact chromatin array.

Furthermore, our experimental strategy provides a time range going from minutes, which is the timescale of gene bursting, to seconds where we can detect the addition of individual coat protein on mRNA molecules following PolII elongation [38]. This approach separates elongation from other kinetic processes taking place on longer timescales, such as splicing or mRNA 3' end processing. Fluorescence correlation analysis was recently employed to study transcription kinetics of individual long MDN1 genes in yeast using the MS2 analog system PP7, and an analytical polynomial-exponential model to fit the fluorescence ACF and extract PolII elongation rates was recently proposed [26]. However, the main assumption used in the reported study, namely that the cassette containing the PP7 sequence is experimentally placed either at the beginning or at the end of a long gene, is not applicable in general, and in particular to U2OS 263 cells, where more than 1 kb of DNA follows the MS2 cassette. Furthermore, the presence of multiple timescales in the intensity trace poses a limit to the applicability of conventional fluorescence correlation analysis.

Instead, we analyzed here the continuous time series using an original approach based on the use of the phasors of the intensity ACF function to fingerprint PolII elongation kinetics. The rates of elongation that we have obtained are in excellent agreement with elongation rates recently measured using biochemical assays. We also confirmed the previously reported large variability in elongation speeds, but this time in the same cell and at loci that are in close proximity (~1 μm). In particular, we observed, as reported in **Figure 3a,** that different cells display a different degree of variance in the observed PolII elongation values but surprisingly, a sizable contribution to this variability arises from differences between the elongation rates of PolII on distinct petals within the same array (**Figure 3b)**. Although the relationship between expression level and copy number of multimerized transgene insertions has been addressed in the past [21], transcription from repetitive sequences is not yet fully understood. Chromosome position effects have been invoked to explain the absence of a linear dependence of transcript to copy number for transgene insertions originating from plasmids in the absence of insulator sequences. Our method allowed us observing that multiple copies of the same gene within the tandem repeats of the U2OS 263 array display a markedly different kinetic behavior, in a context that allows ruling out extrinsic sources of variance, such as cell to cell differences, cell-wide abundance of transcription factors or PolII.

We then asked the question whether there is any correlation between the observed variability of transcription kinetics and the local chromatin mobility, as reflected by the petal angular trajectories. Although investigations on the dynamics of fluorescently labeled loci date back to the 1997 work by Sedat and coworkers[39], only a handful of recent papers have addressed the relationship between transcription and chromatin mobility (reviewed by Dion and Gasser[6]). Notably, Dion *et al*[7] did not observe a consistent correlation between enhanced chromatin movement and transcription activation in yeast. Although mammalian cell lines such as the U2OS 263 systems have been available for over ten years, only recently Khanna *et al*[8] have reported a correlation between HSP70 long range motion and subsequent transcriptional activation. Our analysis of the MSD of the petals trajectories revealed the presence of an active molecular mechanism that, together with diffusion, drives their motion, supporting the notion that the petals originate from the transcription of chromatin loci looping out of the array. Petals that explore a larger area on the surface of the array are also those that display a larger average elongation rate (**Figure 4c**). Taken together with the observed reduction of petal oscillations upon energy depletion (**Supplementary Figure 9**) these measurements point to the role that active, molecular processes must play in determining the motion of the petals. This is confirmed by the cross-correlation between transcriptional activity and angular trajectory of the petals reported in **Figure 4f**, which provides an immediate evidence of a non-diffusive angular motion of an active locus during transcription.

In summary, we introduced a method that allowed monitoring at high spatial and temporal resolution both the kinetics and the physical displacement of multiple transcription sites simultaneously within a model cell line. As illustrated in **Figure 5** our observations demonstrated that elongation kinetics between multimerized copies of a transgene display a large intracellular variance, ultimately affected by factors dependent upon the local microenvironment surrounding otherwise identical genes, rather than being modulated by a more general cell to cell variability. These factors may include the local chromatin mobility and conformation, tuning the accessibility of each locus to transcription factors, PolII and other proteins involved in the transcription process. We also observed that transcription sites that can explore a larger region of space within the nucleus display the higher PolII elongation rates, and moreover their motion occurs in synchrony (cross-correlated) to the transcriptional activity.

Future experiments should compare the transcription kinetics of tandem repeats, which we explored here, to those of individual endogenous loci. In particular, genome editing techniques such as CRISPR/Cas9 will enable not only to fluorescently tag endogenous loci[40], but also to label individual mRNAs[41] and insert fluorescent reporters allowing monitoring the variance of transcription rates and downstream expression at the single gene, single cell level.

Further experiments will be also necessary to elucidate the molecular link between transcription and the movements of the active loci, in order to determine whether these movements are originated by the direct action of PolII or by downstream steps in the transcription chain, such as splicing and mRNA 3' end processing. The current implementation of orbital tracking that allowed us monitoring transcription kinetics in vivo, is amenable to be extended to other high temporal resolution studies of chromatin related processes that involve the in-vivo activity of an enzyme, such as monitoring binding and unbinding rates of transcription factors and other chromatin binding proteins, such as those involved in DNA damage response.

**Methods:**

**Microscopy Setup and Tracking Parameters**

Microscopy experiments were performed on a Zeiss Axiovert 135 microscope frame, modified for orbital particle tracking as discussed in the Supplementary Materials and illustrated in **Supplementary Figure 10.** Using 2-photon excitation from a MaiTai Ti:Sa (Spectra Physics) pulsed IR laser source we employed a single wavelength to track the fluorescence center of mass of the chromatin gene array as well as to excite the EGFP fused to the MS2 coat protein that decorates each newly synthesized mRNA. This strategy frees us from any requirement of alignment of the excitation beams. Excitation at 910 nm corresponds to the peak absorbance of the 2-photon spectrum of EGFP and is also able to excite, although with a 10-fold lower efficiency, the mCherry-labeled locus[42].

During 3D particle tracking the laser beam has to be driven in a closed orbit around the particle. The voltage applied to the scanners was such as to yield a pixel size of 65 nm. A typical radius for the circular orbit was set at 10 pixels. The half-distance between the upper and lower orbit was set at 15 pixels, and we typically performed four orbits above and four orbits below the particle before calculating its $x,y,z$ position using the angular and axial modulation values. Dwell time per pixel was 256 μs, yielding a total time per orbit of 16.384 ms. Output powers below 2mW at the sample plane were employed to minimize phototoxic effects and photo-damage to the living cells.

**Cell culture and transfection**

U2OS 263 cells (a kind gift of Dr. Xavier Darzacq) were cultured in low glucose DMEM (Life Technologies) supplemented with 10% Tet Approved FBS (ClonTech) and 1 % Pen/Strep. Cells were imaged under Leibovitz medium (Life Technologies) supplemented with 10% Tet approved FBS. Cells were transfected using X-treme gene solution (Roche), using an X-treme gene to DNA ratio of 8 µl: 2 µg according to the manufacturer's instructions and after testing for optimal transfection conditions. Cells were imaged 12h to 48 h after transfection. Transcription was induced perfusing the cell wells with 10 µg/ml Doxycycline (Sigma) solution. pTet-On vector was purchased from ClonTech, MS2-EGFP plasmid was obtained from Addgene (plasmid 27121) and mCherry-Lac Repressor was also obtained from Addgene (plasmid 18985). ActinomycinD (Sigma) was used at a concentration of 10 µg/ml. Transfection was also performed using a Gibco Cell Porator Electroporation system, according to the manufacturer's instructions and with the following parameters (180 V, 1180 µF, low Ohm), obtaining similar results to chemical transfection.

**Phasor Analysis**

The phasor analysis of autocorrelation curves as the one displayed in **Supplementary Figure 4c** is performed according to the standard equations introduced in the past for lifetime or spectral analysis[28][29][31]. For each autocorrelation curve described by the function $F(t)$ a point on the unit circle in the complex plane is calculated using the $S$ and $G$ coordinates, according to **Equation 1**

$$S(\omega_1) = \frac{\int F(t)\sin(\omega_1 t)\,dt}{\int F(t)\,dt}$$

$$G(\omega_1) = \frac{\int F(t)\cos(\omega_1 t)\,dt}{\int F(t)\,dt}$$

**Equation 1**

$\omega = \frac{1}{2N\Delta t}$ is the fundamental frequency and is determined by the temporal resolution $\Delta t$ of our measurement and by the time bin used to calculate the ACF, typically 256 s yielding $N=1024$. The angle $\phi$ of each *phasor* component is given by: $\varphi = a\tan(\frac{S}{G})$. All data analysis was performed using custom written IgorPro routines and SimFCS (Globals Software).

**Acknowledgments**

Work supported in part by NIH grants P50 GM076516 and P41 GM103540. We would like to thank all the LFD staff for support, in particular A. Dvornikov for support with mechanical machining and H. Chen for useful discussion on electronics and data acquisition. We would like to thank Dr. Kyoko Yokomori and Dr. Marco Marcia for critical comments on the ms. We would like to acknowledge X. Darzacq for kindly providing us with the U2OS 263 cell line.

**Author Contributions**

E.G. and P.A. designed the experiments. P.A. performed the experiments. P. A. and E.G. analyzed the data. P.A. and E.G. wrote the manuscript.

**Competing financial interest**

The authors declare no competing financial interests.

# Figure Legends

**Figure 1: Imaging mRNA synthesis distinct regions of an activated transgene array at high spatio-temporal resolution. a)** Multicolor Laser Scanning micrograph of the chromatin array (LacI-mCherry: red) and the mRNA (MS2-EGFP: green) in U2OS 263 cells following Dox induction (+30'). Excitation is provided with 561 nm and 488 nm laser. Petals of mRNA synthesis are clearly visible. **b)** Schematics of the experimental configuration for 3D orbital tracking using a 2-photon laser-scanning microscope. The difference in the fluorescence intensity collected from a circular orbit above and a circular orbit below the particle determines its axial coordinate ($z$-position) [19]. The intensity profile along each circular orbit is used to find the $x$ and $y$ position of the fluorescence center of mass in the imaging plane, using a Fast Fourier Transform based localization algorithm. The particle can be followed using a feedback loop that re-centers the orbit on the particles at each cycle. One orbit is performed every 32 ms and a full 3D localization cycle happens within eight orbits. **c)** Kymograph of the fluorescence intensity collected along the entire orbit, reflecting the presence of five globular regions or petals. Each line is calculated integrating the fluorescence intensity of 16 orbits, yielding a temporal resolution of 260 ms. The black line highlights the angular trajectory of one of the petals. **d and e)** Detail of a 30 s interval of the fluorescence kymograph, and Gaussian fit of the line intensity profile. The center of the petal is localized with a precision < 6 nm. **f)** Intensity profile measured from the carpet displayed in **e** **g)** Sub μm displacement trajectory of the petal.

**Figure 2: pCF analysis of mRNA molecules leaving the petals and flowing into the nucleoplasm. a)** Schematics of the laser trajectory around the transgene array. The laser PSF performs a trefoil orbit and the lobes of the trefoil reach into the nucleoplasm about 1 μm away from the center of the array. A distance of 12 pixels along the rows of the carpet is used to calculate the pCF carpet. The fluorescence collected at each point along the orbit is cross-correlated to that recovered at a point 12 pixels away (a variable distance given the shape of the orbit). The result of the cross-correlation at each instant of time is the Pair Correlation Function (pCF). **b)** The pCF calculated for each point of the orbit can be used to calculate the pair correlation carpet (pC carpet). Each column of the pC carpet indicates the degree of connection (due to diffusion or flow of fluorescently labeled mRNA) between two positions spaced of 12 pixels along the orbit. Intensity at short times in the 1to2, column indicated diffusion of mRNA molecules from the petals, rather than from the center of the array (no cross-correlation in the 3to5 column). The pCF 4to6 highlights a delay for the mRNA to reach the nucleoplasm from this petal. The mRNAs diffuse freely in the nucleoplasm, as indicated by the 5to7 column, but cannot reach the center of the dense chromatin array (no cross-correlation is observed in the transition 8to9). The amount of delay for the mRNA in reaching the nucleoplasm is again different when looking at the last petal (10to11). The pC carpet was smoothed to reduce noise. **c)** Schematic representation of the pCF at the positions where newly synthesized mRNAs leave the petals. Delays of up to a few seconds in reaching the lobes of the orbit can be observed.

**Figure 3 Measuring heterogeneous PolII elongation kinetics across cells, within cells and within petals a)** PolII elongation rates measured using the Phasor Method in different interphase cells. Letters *A-K* identify 11 different U2OS 263 **cells** containing the transgene array. The scatter plot for each cell is obtained by combining the elongation measurements on multiple petals. The distribution of elongation rates integrated for all the cells is labeled as *ALL*. Symbols indicate previous measurements of PolII elongation using the MS2 system: *Diamond* Maiuri *et al*; *Cross* Darzacq *et al*; *Dot* Hocine *et al*; *Star* Boireau *et al*; *Circle-dot* Larson *et al*; *Square* Yunger *et al*. Blue: mammalian cell lines. Green: yeast. **b)** Scatter plot of PolII elongation rates measured **in the petals** inside the cells reported in panel a. Mean Values and Standard Error of the Mean are superimposed. **c)** Fluorescence kymograph of a petal observed to display increasing MS2-EGFP intensity over a period of approximately 35 minutes after Dox induction, indicative of activation. **d)** (top) Intensity counts for the kymograph in **c** and (bottom) corresponding elongation rates as a function of time. **e)** Intensity carpet of a petal upon treatment with 10 μg/ml AD. **f)** (top) Intensity counts for the kymograph in **e** and (bottom) corresponding elongation rates as a function of time.

**Figure 4: Correlation between mRNA transcription kinetics and petal mobility. a)** The petals display sizable motion along the surface of the transgene array (left). Localized positions (using the Particle Tracker Mosaic [43] ImageJ Plugin) of the petals over time reveal that this motion is predominantly angular (indicated as $\phi$ in the figure). **b)** MSD analysis of the angular ($\phi$) displacement of the petals. Black dashed lines: experimental MSD of the petals. Orange line: average of experimental MSD. Guide for the eye: MSD$\propto t^{0.75}$ for time scales <5-10 s (dotted orange line) and MSD$\propto t^{0.25-0.5}$ (dashed-dotted blue and green lines) between 10-100 s. **c)** Scatter plot of the confinement radius $R_{max}$ (calculated extrapolating the 10-100s MSD power law fits to 512 s) against the elongation rate measured for each petal. Superposed is a linear regression. Pearson correlation coefficient is R=+0.52 with a p-value=0.008. **d)** Simulated trace of intensity fluctuations (bottom) and angular ($\phi$) displacement (top) of a petal for counterclockwise (blue trace), clockwise (green trace) and bidirectional (black trace) motion. **e)** Calculated cross-correlation for each of the three cases. A peak in the cross-correlation function is present only if there is a predominant directional motion. **f)** Binary map of cross-correlation of the fluorescence intensity and the angular ($\phi$) displacement of the petals, calculated in 128 s time bins. Each line corresponds to a different experiment, and the sign of the cross-correlation amplitude is arranged in order to display negative values first. The average cross-correlation function measured for each petal is displayed ($\pm 1$) only if above the 99% confidence band for pure diffusion.

**Figure 5 Model of transcription and mRNA export from multimer copies of the same locus within a transgene array.** Transcription is observed to occur only on the surface of the

transgene array, in distinct and mobile structures named petals. Each petal undergoes different transcription kinetics (elongation rates $k_1$, $k_2$) and releases mRNA molecules in the nucleoplasm with characteristic times. The most mobile petals display higher elongation kinetics and each petal moves along a preferential direction as the amount of labeled mRNA increases, and recoils after a 10-100s delay.

# Figures

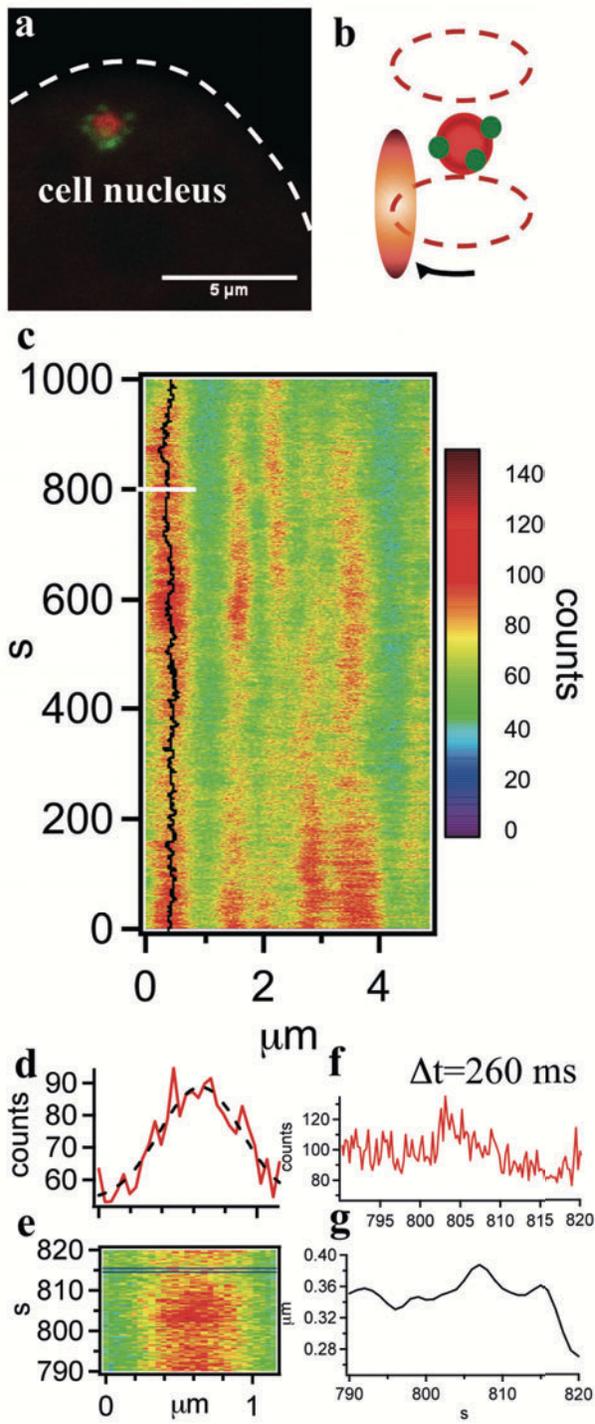

Figure 1

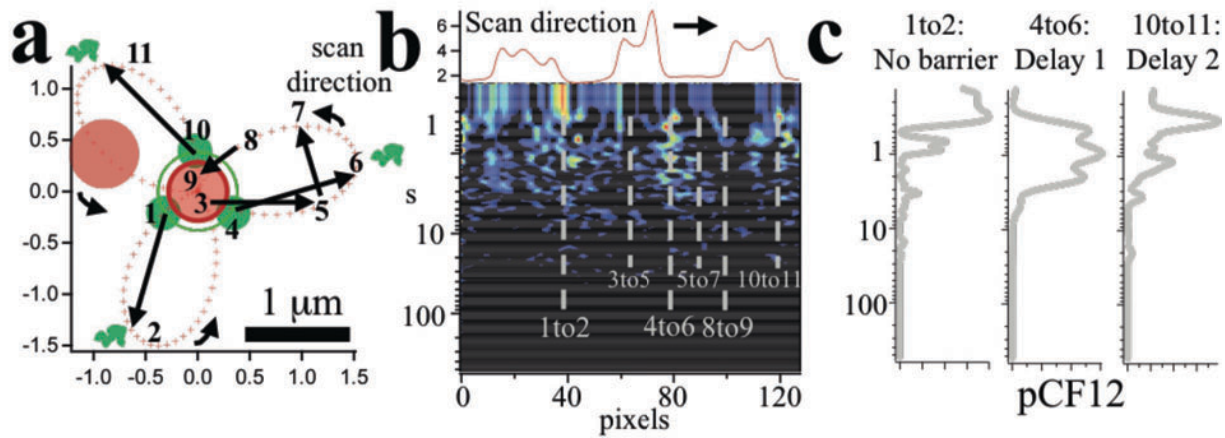

Figure 2

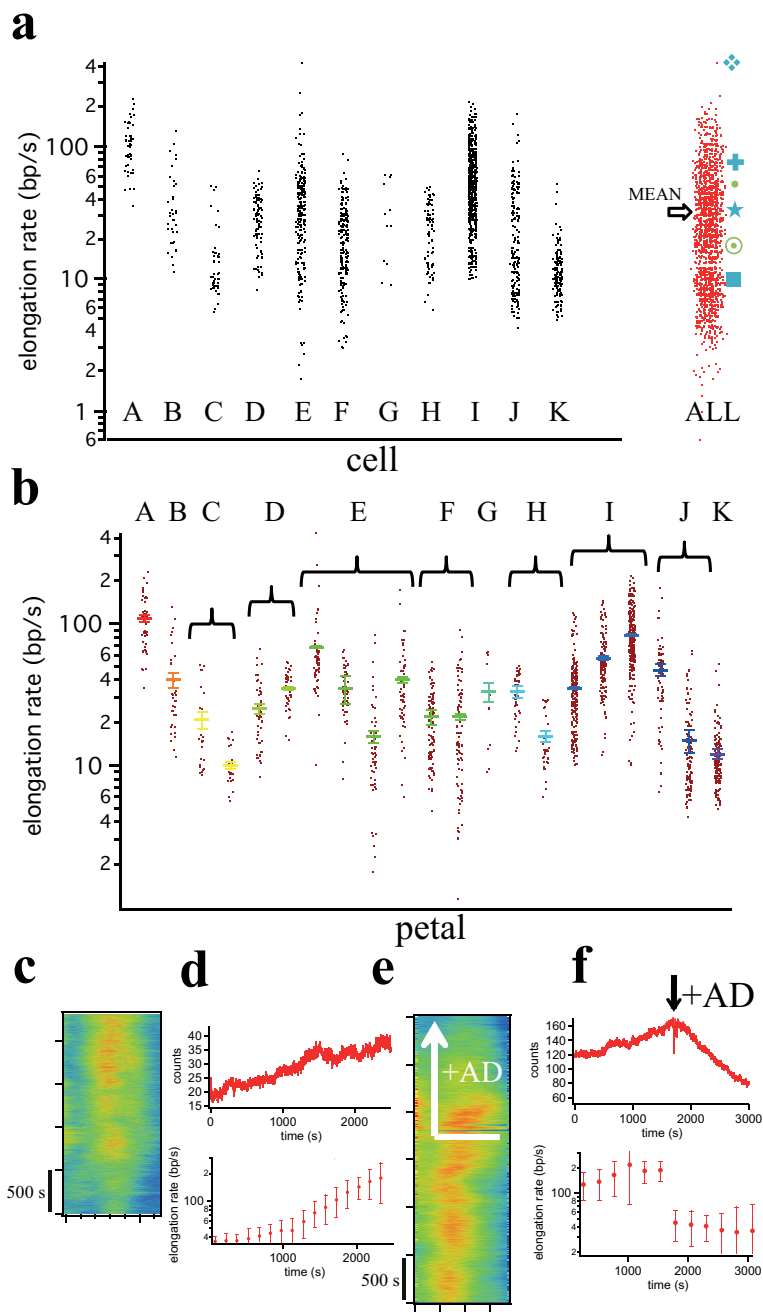

**a**

**b**

**c**  **d**  **e**  **f**

Figure 3

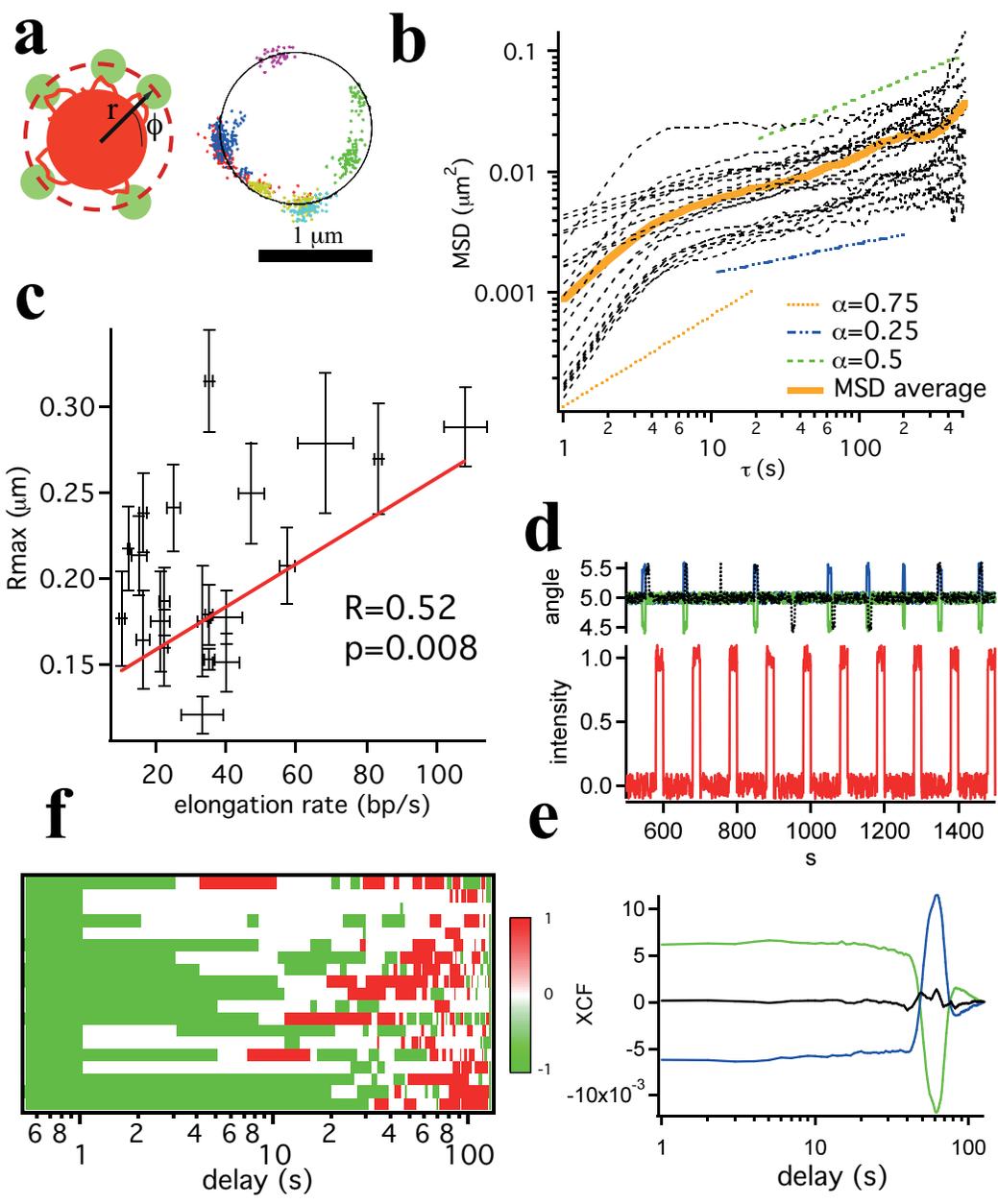

Figure 4

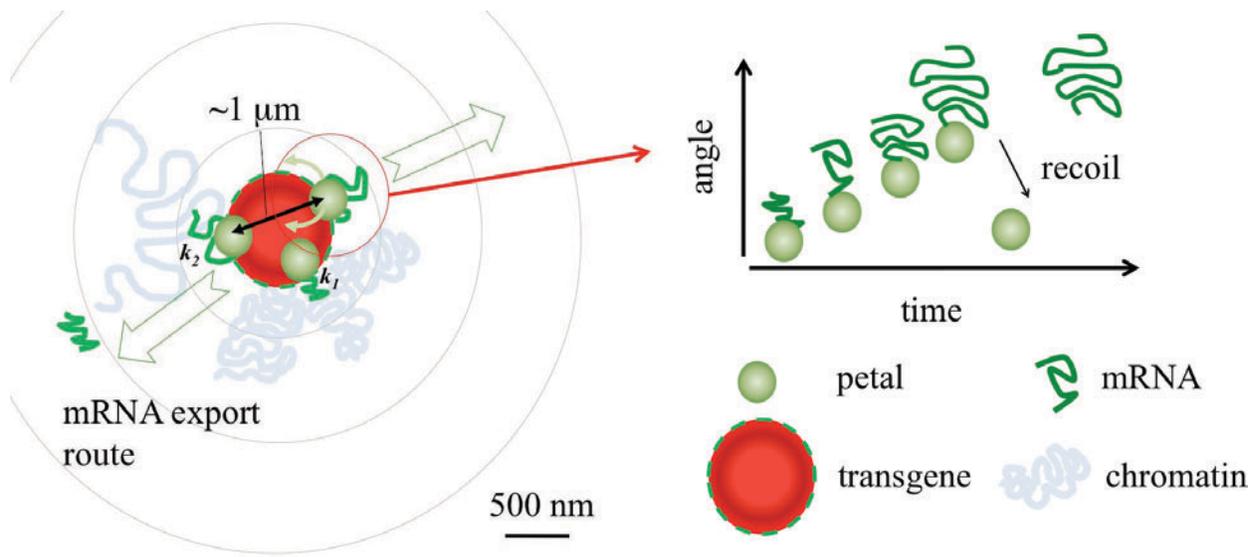

Figure 5

# Supplementary Materials

# Single cell visualization of transcription kinetics variance of highly mobile identical genes using 3D nanoimaging

Paolo Annibale, Enrico Gratton

**Supplementary Figures**

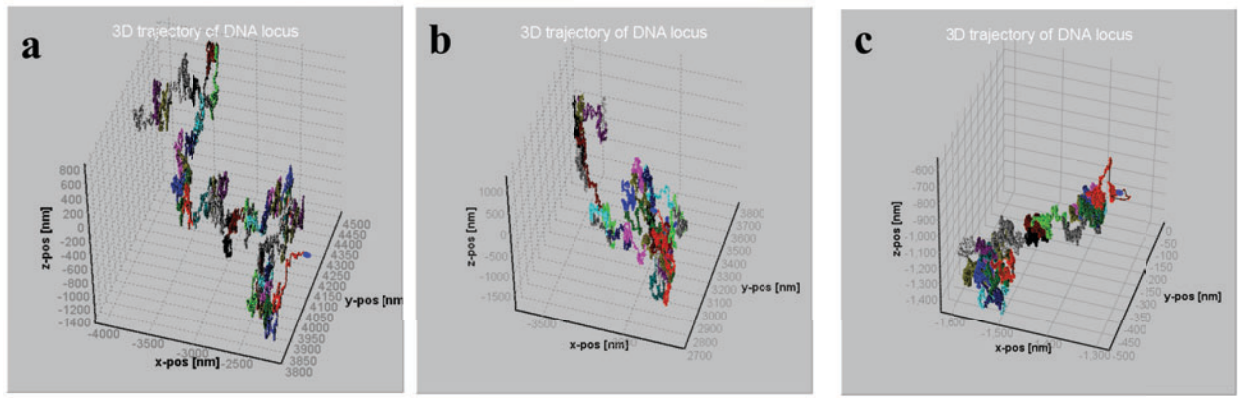

**Supplementary Figure 1. Reconstructed 3D trajectory of the fluorescence center of mass of the transgene array within the nucleus of U2OS 263 cells in three representative cases following Doxycycline induction.** a) The array displays a maximal displacement of 2 µm along the x direction.  b) In this trajectory the maximum displacement occurs along the z direction, for a total of about 2.5 µm. c) Trajectory of a transgene array following 30' minutes incubation with 10 µM Sodium Azide and 50 µM 2-Deoxyglucose. The maximal displacement has been reduced to sub-µm size.

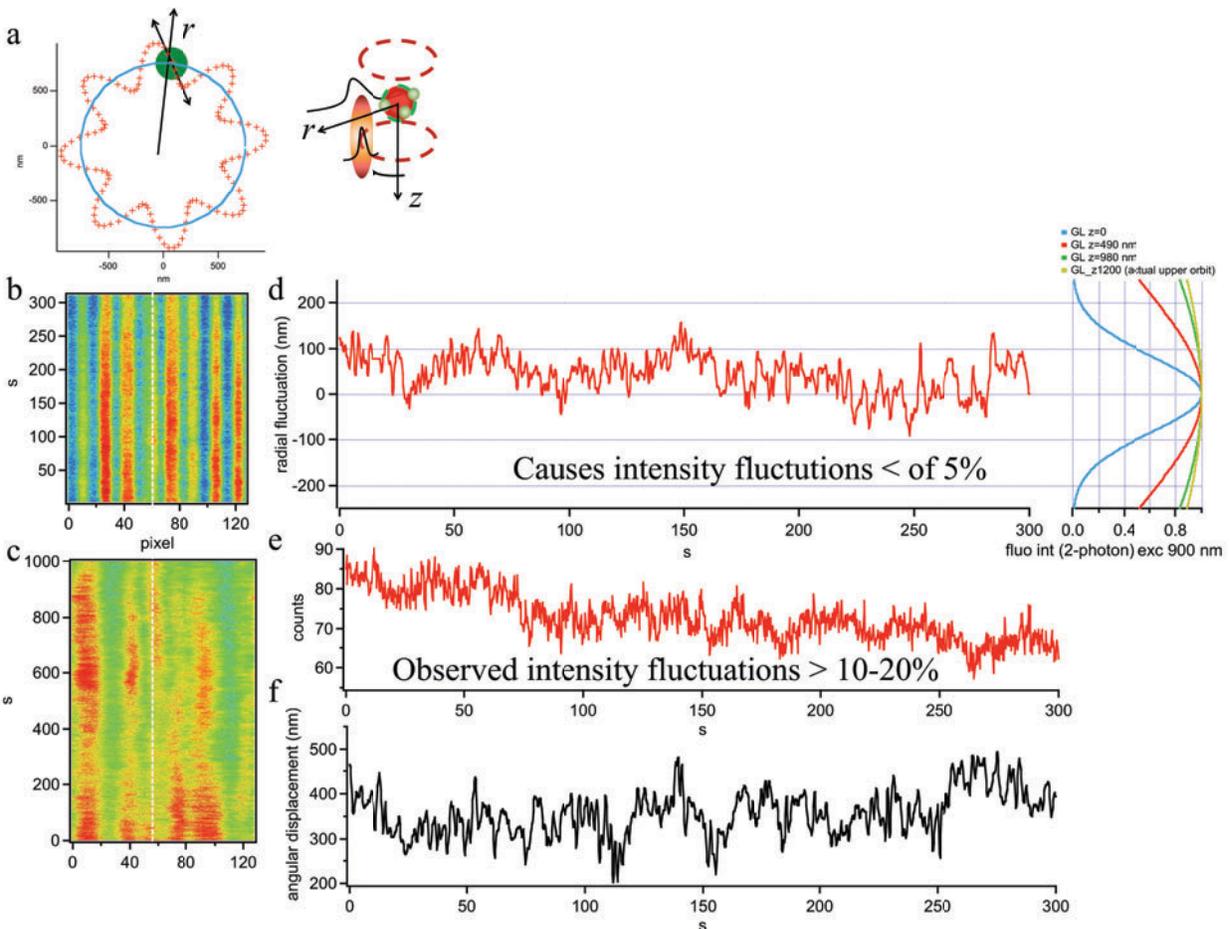

**Supplementary Figure 2 Effect of the spatial fluctuations of the petals on the fluorescence intensity traces collected performing orbital tracking.** The fluctuations of the petals are predominantly angular. The contribution of radial fluctuations can be examined by switching from a circular orbit to a Rosetta shape. a) A Rosetta orbit intersecting a petal allows monitoring the radial component of the displacement of the petals. In 3D orbital tracking geometry, only the distal portion of the PSF along the z-axis intersects the petals. b) Intensity carpet collected along a Rosetta orbit and c) intensity carpet collected, shortly after, along a circular orbit surrounding the same transgene array. d) The radial displacement of a representative petal (in nm) is displayed together with the profile of the Gaussian-Lorenzian Point Spread Function of the 2-

Photon microscope for an excitation wavelength of 900 nm. According to the relative z-distance of the orbit with respect to the plane of the petals, the PSF broadens along the radial direction. For the typical distance of the upper/lower orbits from the center of the transgene array of approximately 1 µm, the PSF is so broad that the intensity fluctuations caused by radial oscillations would contribute less than 5% of the total intensity. e) Intensity fluctuations collected during a circular scan on the same petal as in d. f) Angular displacement of the petal along the circular orbit.

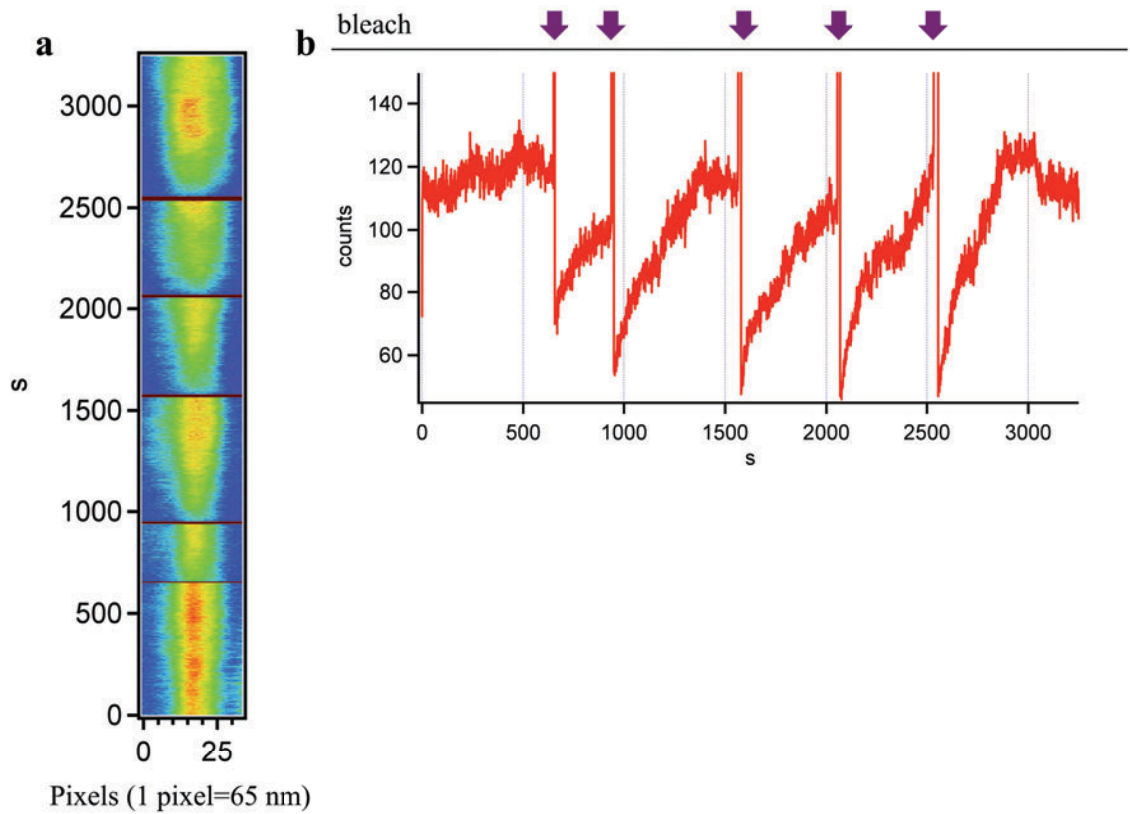

**Supplementary Figure 3. Tracking-FRAP of an active petal** a) Intensity carpet and average fluorescence trajectory of a *petal* subjected to short pulses of intense excitation light to cause photobleaching of the fluorescently labeled mRNA molecules. The unit of the lower axis is pixels along the orbit. b) Recovery of the fluorescence intensity at each cycle confirms that the petal is a genetic locus undergoing active transcription.

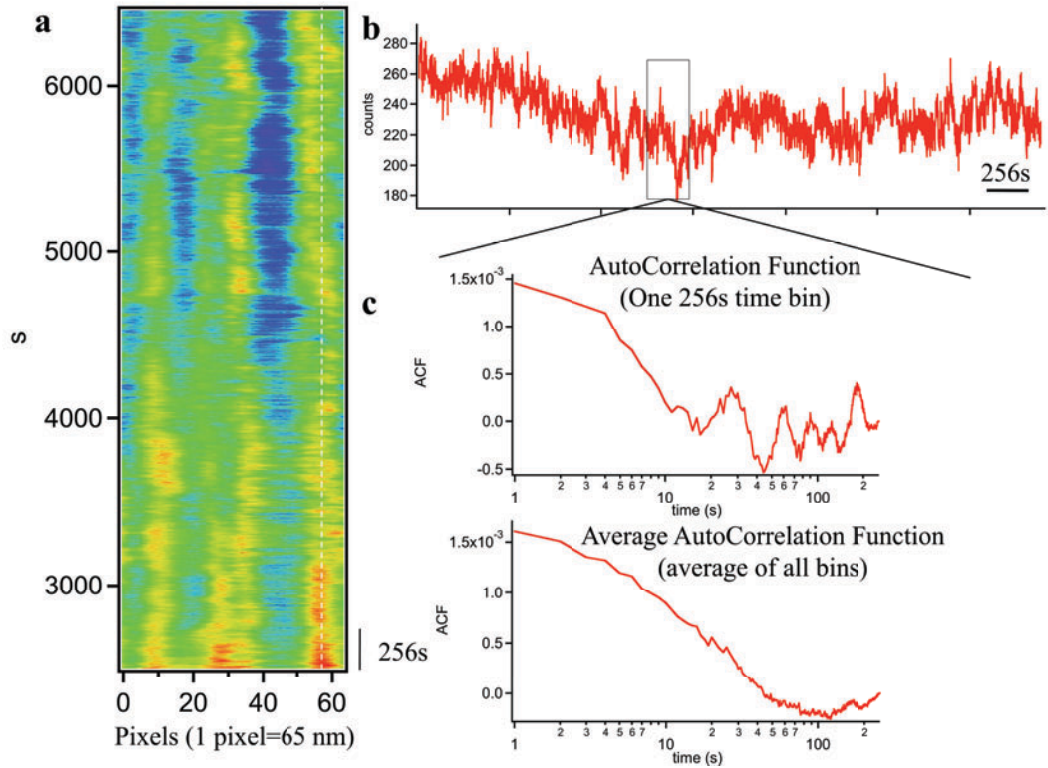

**Supplementary Figure 4. Autocorrelation analysis of the petal carpets** a) Intensity carpet and

b) Fluorescence trajectory. c) AutoCorrelation (ACF) function calculated starting from short time

bins. In this example time bins of 1024 points (256 s) are used. The global autocorrelation function of an active petal is calculated by averaging all the ACFs of individual bins.

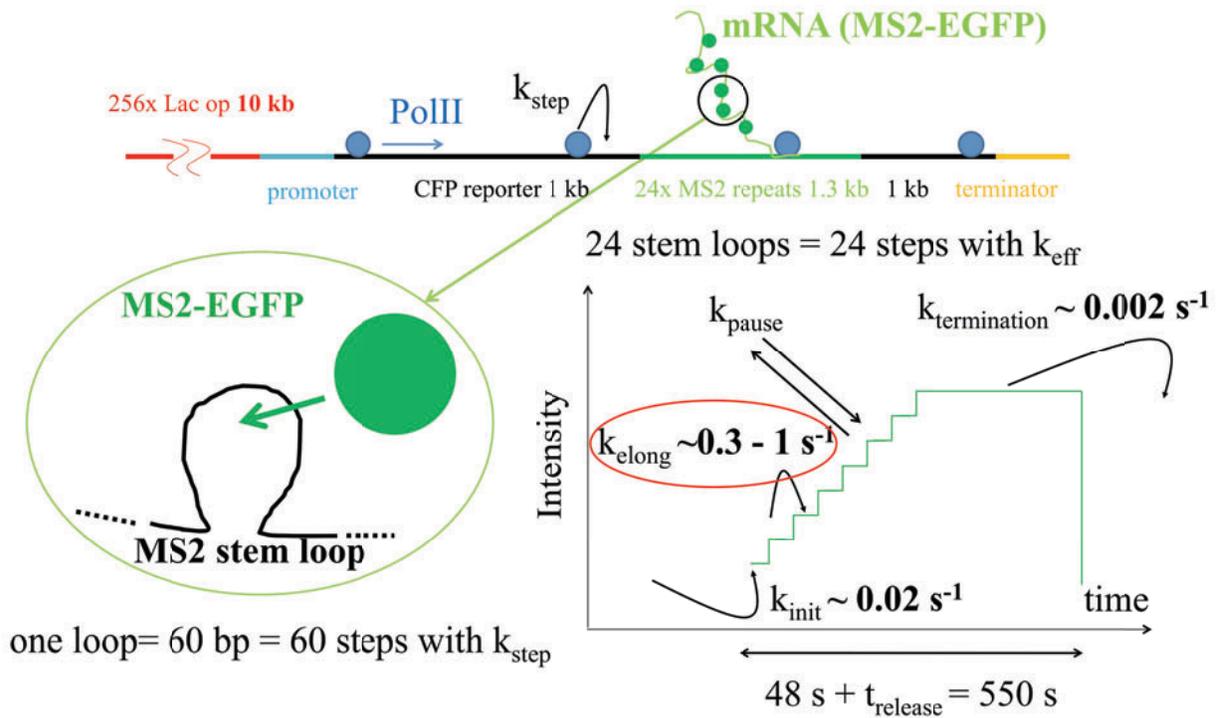

**Supplementary Figure 5. Kinetic model for the analysis of MS2-mRNA fluorescence traces.**
**a)** Diagram of an individual repeat within the U2OS transgene array, with annotated length of each relevant portion (reporter gene, MS2 cassette, downstream sequence). b) The periodicity of the MS2 stem loops is about 60 bp, and the MS2-GFP construct binds as a dimer. c) Changes of fluorescence intensity on the gene due to the synthesis of the mRNA molecule. Kinetic rates are taken from the work of Darzacq *et al* [1].

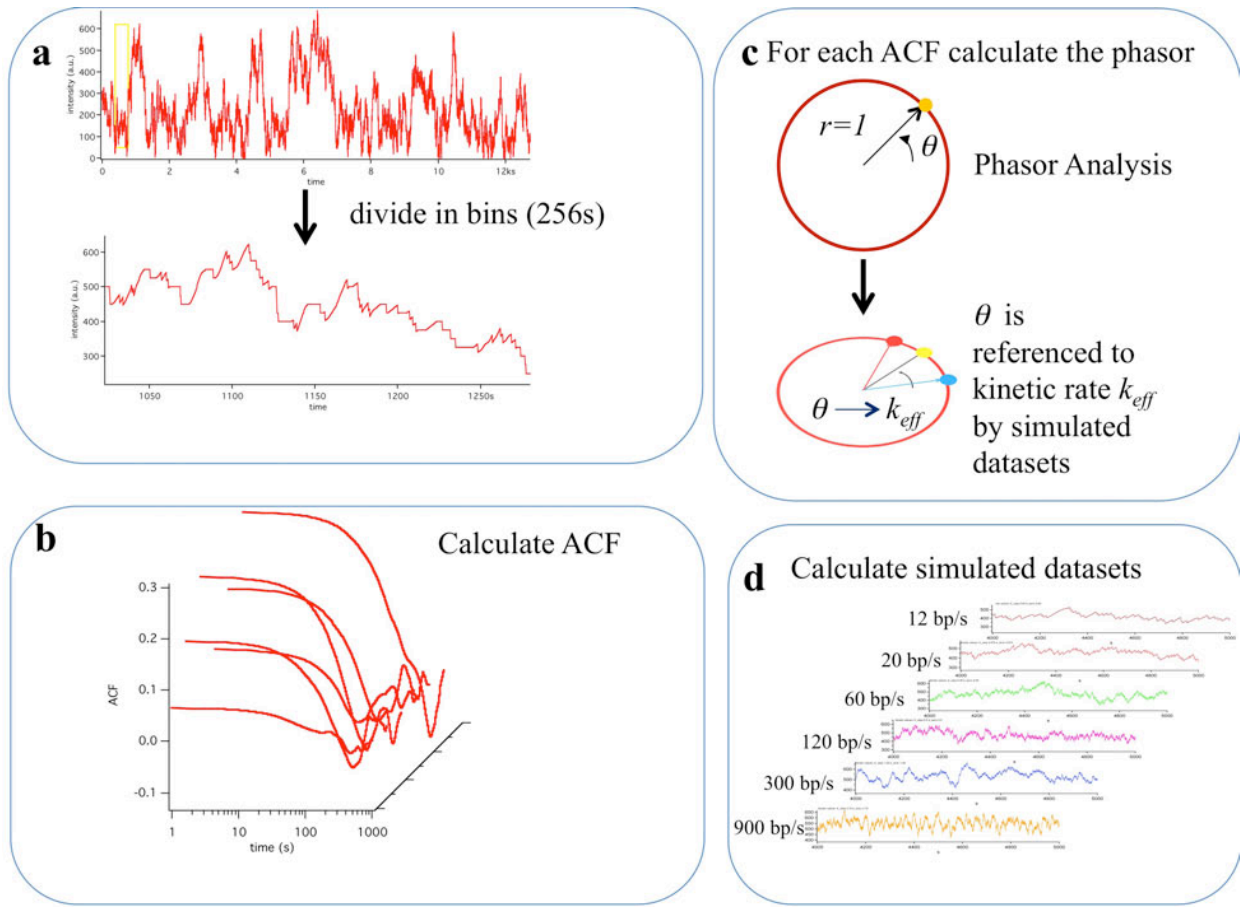

**Supplementary Figure 6. Phasor analysis of the experimental data**. a) The fluorescence trajectory (here a simulated dataset) of an individual petal is divided in time bins (N=1024 pts = 256 s) and the b) Autocorrelation Function (ACF) of each is calculated. c) The phasor analysis of the autocorrelation function converts the kinetic information within the ACF into an angle in the complex plane. The experimental points can be referenced using simulated datasets. d) Simulated trajectories are calculated for varying kinetic parameters (elongation rate, termination rate).

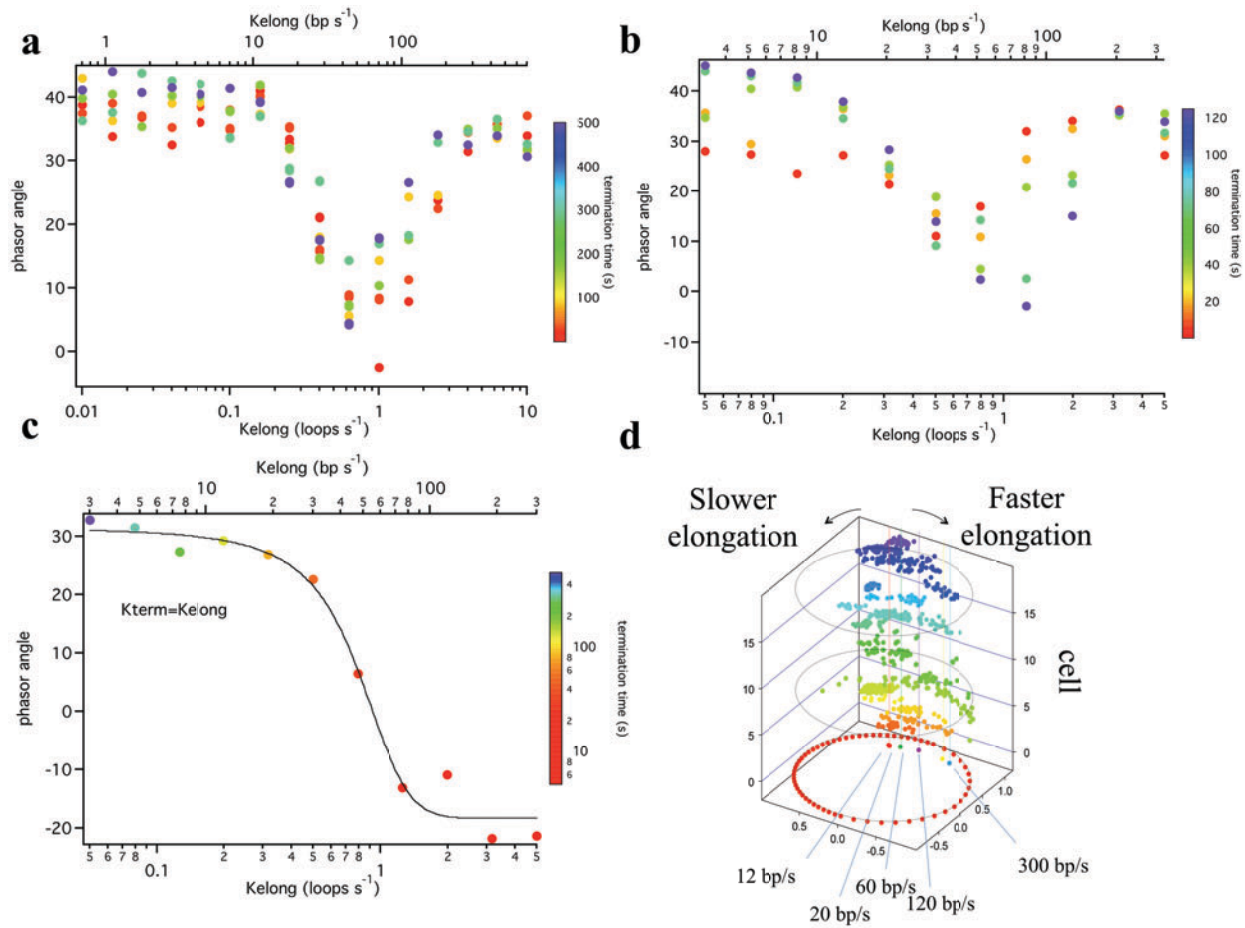

**Supplementary Figure 7 Phasor analysis of simulated and experimental datasets.** a) Phasor angle (degrees) as a function of the elongation rate ($k_{elong}$) for varying average termination rates, assuming an exponential distribution of termination rate. b) Phasor angle as a function of the elongation rate ($k_{term}$) for varying termination rates assuming a fixed termination time. c) Dependence of the phasor angle upon the elongation rate assuming $k_{elong}=k_{term}$, viz that PolII runs through the last 60 bp of the gene at the same speed it does on the MS2 cassette and is then instantly released. d) 3D scatter plot of experimental phasor points calculated from the fluorescence kymographs of petals in different cells. Reference phasor points are indicated at the bottom of the graph, according to the calibration curve in panel c.

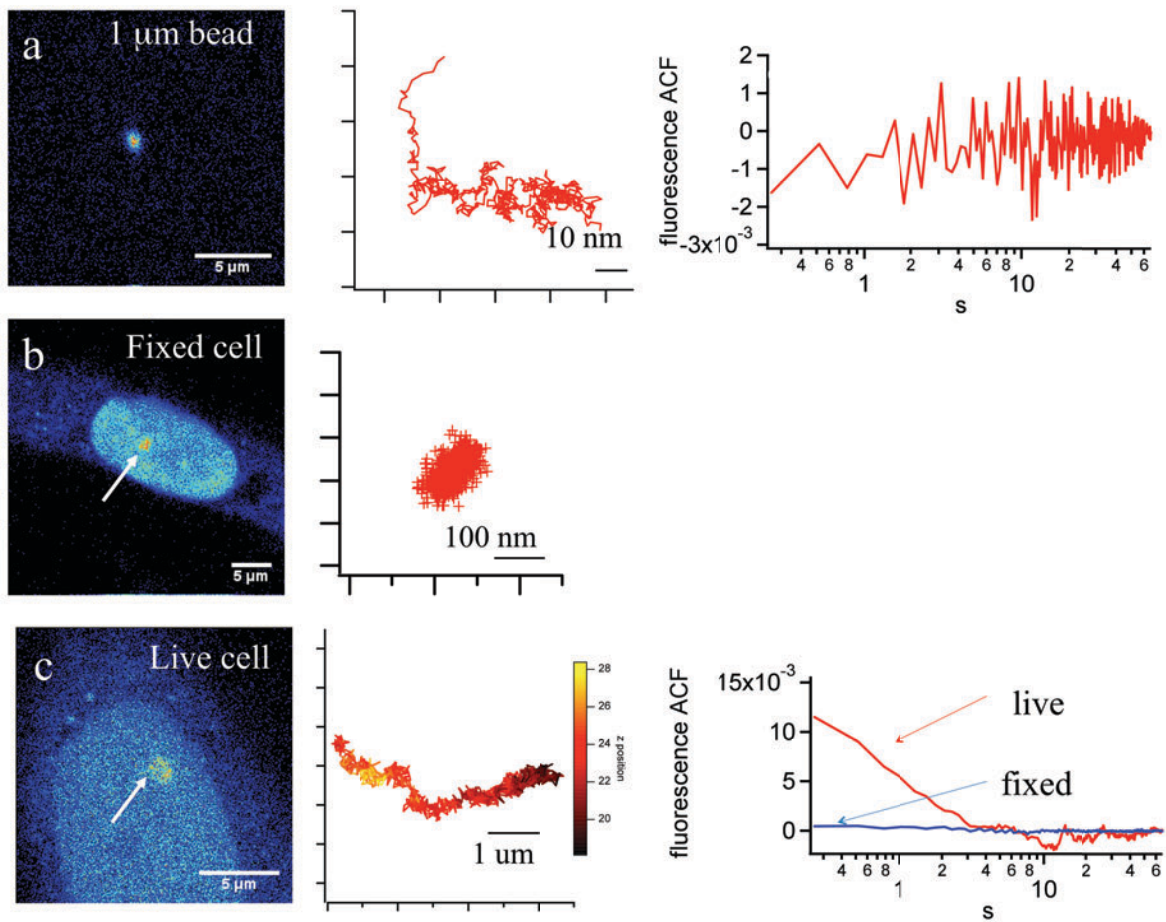

**Supplementary Figure 8 Control experiments.** a) Tracking of an immobile 1 µm bead on a glass surface. Raster scan image (left), x-y projection of the 3D trajectory (center) and autocorrelation function of the fluorescence intensity (right). b) Tracking of an inducted array within a fixed cell. Raster scan image (left) and x-y projection of the 3D trajectory (right). c) Tracking of an inducted array within a living cell. Raster Scan image (left), x-y projection of the 3D trajectory (center) and Autocorrelation function of the fluorescence intensity collected from the MS2-EGFP constructs compared to the ACF collected on a fixed cell (right).

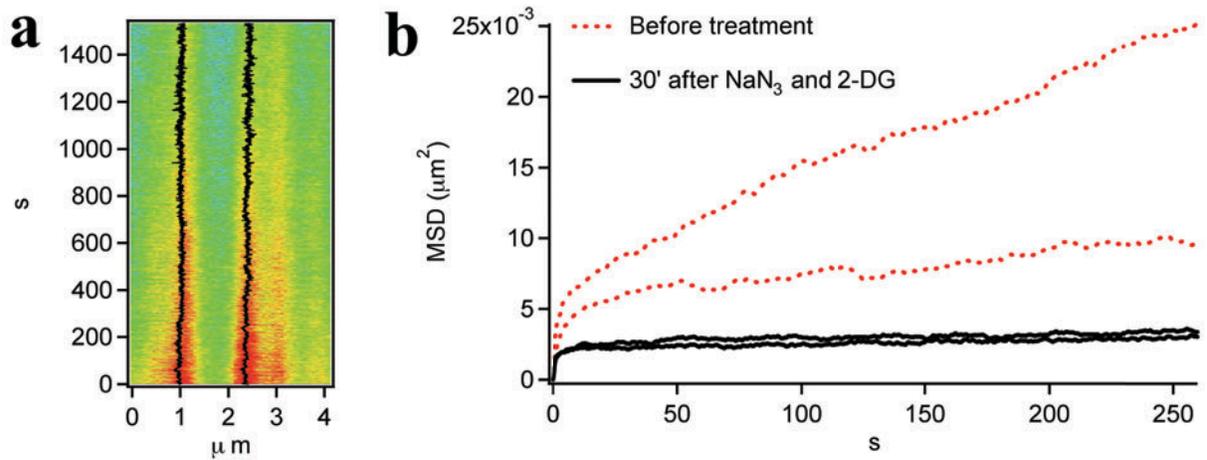

**Supplementary Figure 9 ATP depletion treatment** a) Representative intensity carpet displaying the fluorescence of mRNA petals surrounding the transgene array in a cell treated with 10 mM Sodium Azide and 50 mM 2-Deoxyglucose for 45'. Angular fluctuations in the petals motion are effectively suppressed. The progressive decrease in fluorescence intensity of each petal is due to photobleaching of the MS2-EGFP, not compensated by the synthesis of new mRNAs. b) Comparison of the Mean Squared Displacement (linear scale) calculated on two petal trajectories before and two trajectories after ATP depletion. ATP depleted trajectories display a minimal residual motion, indicating a corralling size $(MSD)^{0.5}$ of about 95 nm, corresponding to 1.5 times the pixel size of the measurement.

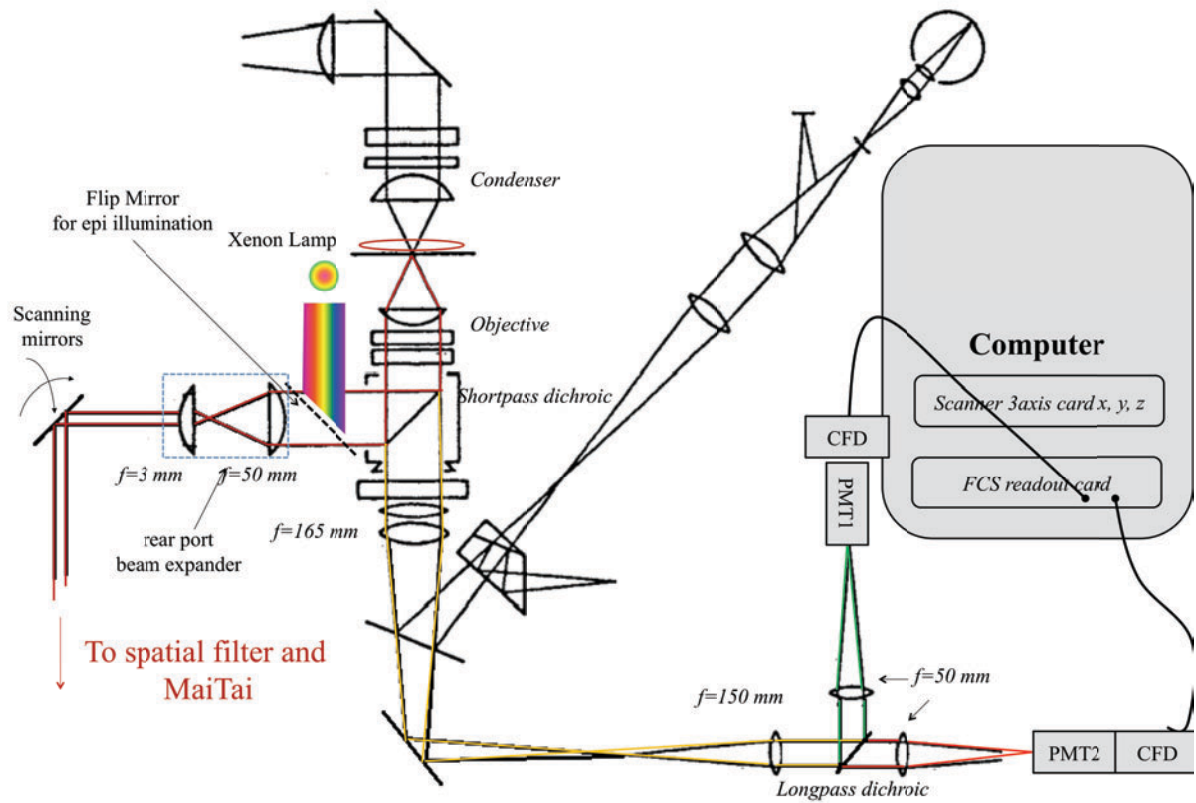

**Supplementary Figure 10 Schematics of the tracking microscope**. The expanded and collimated infrared beam from an Ti:Sa Laser (MaiTai, Spectra Physics) is coupled into the rear port of a Zeiss Axiovert 135 inverted microscope. A pair of scanning mirrors placed in the focus of a beam expander allows steering the beam in the sample plane. A 60X 1.2 NA Olympus water objective is used. The fluorescence is collected through a dichroic short-pass filter and split between two detectors using an additional long-pass dichroic mirror. The signal from the photomultiplier tubes (PMT) is amplified, discriminated (CFD) and the digital output is read out using a fast FCS data acquisition card.

**Supplementary Materials and Discussion**

*Experimental setup*

The excitation laser beam was provided by a MaiTai Ti:Sa (Spectra Physics) laser source, providing pulsed IR excitation tunable in the range between 690 and 1040 nm. The power of the excitation beam is controlled using an AcoustoOpic Modulator (AOM) MT110 (AA Optoelectronics, Orsay, France). The first order beam output from the AOM is beam expanded to approximately 0.5 inches and spatially filtered through a 15 μm pinhole (Thorlabs).

Microscopy experiments were performed on a Zeiss Axiovert 135 microscope frame, modified for orbital particle tracking as illustrated in **Supplementary Figure 10**. The rear port was fitted with a customized aluminum lens holder to accommodate a short focal distance lens (f=3 mm, a microscope eyepiece) and a collimating f=50 mm lens mounted on a 1 inch cage system (Thorlabs). The two lenses act as an effective beam expander able to overfill the back focal plane (BFP) of the microscope objective (Olympus 60X, 1.2 NA Water). The beam is reflected into the objective BFP after being reflected by a 680 nm short pass dichroic mirror (Semrock). A 1-inch flip mirror (Thorlabs) was inserted after the collimating lens to switch from laser excitation to wide-field excitation using the slightly defocused beam originating from a Xenon Lamp.

A pair of Galvanometer Scanning Mirrors (Cambridge Technologies) was installed in the focal point of the f=3 mm lens (a plane in the Fourier Space with respect to the sample plane, hence a change in angle reflects in a change in lateral position in the sample plane) in order to scan the beam in the sample plane. The scanners were mounted and adjusted on the optical axis of the rear port beam expander using a pair of two inches steel posts that were then clamped to the optical table.

The emitted fluorescence is collected through the bottom port of the Axiovert 135 frame, and re-collimated using a 150 mm achromatic doublet (Thorlabs). The collimated fluorescence is split through a 565 longpass dichroic mirror (Chroma) and bandpass filtered (510-20 nm and 610-50 nm, Chroma) to be directed to two Hamamatsu H7422P-40 Photomultiplier Tubes. Detection is performed in photon counting mode, and the signal output from each Photomultiplier tube is electronically amplified (ISS 40493) and discriminated using a Constant Fraction Discriminator model 6915 (Phillips Scientific). The TTL output of the discriminator is supplied to a fast readout card FCS 1.1 PCI Card (ISS) with a maximal acquisition frequency of 500,000 Hz.

XY translation of the sample is achieved using an MS2000 stepper motor stage (ASI), with a controller for the coarse axial positioning of the objective turret. Furthermore, the objective is installed on an A260 *Z*-axis piezoelectric nano-positioner stage (ISS).

The voltage to control the Galvo Scanners and the nanopositioner Stage is provided using an ISS 3-axis PCI card. The sample is kept at 37 °C by the use of a PDMI-2 MicroIncubator (Harvard Apparatus), adapted onto the MS2000 stage. The entire data acquisition via the FCS card and the control of the scanning microscope via the 3-axis card is performed using the SimFCS® software, developed by Enrico Gratton at the Laboratory for Fluorescence Dynamics.

*MonteCarlo Simulations of Pol II motion along the gene and generation of the fluorescence trajectory*

We run simulations of fluorescence trajectories of the MS2-EGFP signal upon a given set of kinetic parameters for PolII elongation on our gene (**Supplementary Figure 5**). In our case, the use of simulated datasets is necessary to provide an absolute calibration to the phasor plot values. As discussed in the main text, the phasor plot of the data can tell us that different petals display markedly different elongation kinetics, but cannot provide us with absolute values unless the elongation rate of the PolII were known beforehand. For this reason we had to recur to simulated trajectories for varying values of the elongation rate and the termination rate to calculate the reference phasor points that are displayed in **Supplementary Figure 6d** and used to obtain the kinetic values in reported in **Figure 3**.

The final observable in the simulated experiment is the fluorescence intensity, increasing from the addition of MS2 subunits to the nascent mRNA and decreasing by the detachment of complete mRNA molecules. Each mRNA molecules carries 24x MS2 stem loops, each of them can be bound by an MS2-EGFP dimer. For simplicity we have assumed that all the loops are occupied (although we are well aware that Wu et al. [2] demonstrated that for the MS2 system only half of the sites end up being occupied, even at high MS2-EGFP monomer concentrations) and that the dissociation constant for MS2-EGFP to the mRNA stem loops is negligible. In each gene of the trans-gene array of U2OS 263 the 24x MS2 cassette is about 1.5 kb long and is followed by approximately another 1 kb of downstream gene before the PolyA site.

We model the gene as a sequence of units, each 60 bp long, corresponding to the approximate length of an MS2 stem loop. The first 24 of these units are active, meaning that once occupied by a PolII one unit of fluorescence is added to the total fluorescence at that time point, while the remaining 16 do not give rise to any fluorescence. Once a PolII reaches the end of a gene the total fluorescence is diminished of 24 units. For modeling purpose we neglect the portion of the gene before the MS2 cassette, since it does not give rise to any fluorescence signal. A first order kinetics is used to model the stepping forward of PolII from one of these units to the next one. This is along the lines of what reported in Larson et al. [3] but differs significantly from other approaches, such as the TransWave model employed by Maiuri et al. [4]. If a PolII occupies a site, upstream PolII enzymes cannot occupy that site, hence, if a PolII pauses, a pile-up of trailing enzymes is formed. An effective kinetic constant, that is an effective elongation rate is used and is a parameter that can be tuned within the model. First order kinetics is also used to model PolII initiation, release from the gene and transition and recovery to a paused state.

In the current report we followed two fundamental assumptions that were previously employed by [3] in deriving an analytical model for autocorrelation of PolII elongating along MDN1 gene in yeast. First, the mRNA release constant was assumed identical to the effective elongation rate. This is a significant assumption, since a slow release of the mRNA from the gene could be the rate-limiting step. However, as illustrated in **Supplementary Figure 7 a-b**, even significant variations of the PolII termination rate do not impair the ability to detect changes in elongation rate in the relevant range 8-80 bp/s. Second, PolII is assumed not to pause while transiting over the MS2 cassette.

Accepting these two assumptions, we are left with only two parameters to set, initiation and effective elongation. We inserted the initiation rate in our model as a constant, using the value reported by Darzacq et al. of 0.0216 /s. Changing the initiation value did not affect significantly the result of our analysis. The elongation rate was then varied between 10 bp/s and 900 bp/s, giving rise to the simulated trajectories reported in **Supplementary Figure 6d**.

It should be further noted that if the assumption by Larson *et al.* [3] that the termination rate and the effective elongation rate are equal does not hold (i.e. $k_r \neq k_s$), then the elegant form of the autocorrelation function for short genes determined by the authors,

$$G(\tau) = \frac{k}{c}\left(\frac{2}{3}\right)\frac{1}{\left(N(N+1)\right)^2}e^{-k\tau}\sum_{n=0}^{N}(N-n)(N-n+1)(2N+n+1)\frac{(k\tau)^n}{n!}$$

**Equation 1**

turns into a considerably more complex expression. For the simpler case of only 5 stem loops, we have obtained an analytical equation for the autocorrelation function (posing the elongation constant $k=k_s$, and defining a termination constant $k_r$):

$$G(t) = \frac{k_i}{6k_r k_s (k_r - k_s)^4}\left\{e^{-k_s t}\left[t^3\left(k_r k_s^7 + k_r^2 k_s^6 - 9k_r^3 k_s^5 + 11k_r^4 k_s^4 - 4k_r^5 k_s^3\right) + t^2\left(12k_r k_s^6 + 12k_r^2 k_s^5 - 93k_r^3 k_s^4 + 102k_r^4 k_s^3 - 33k_r^5 k_s^2\right) + \ldots\right]\right\}$$
$$\ldots + t\left(60k_r k_s^5 + 60k_r^2 k_s^4 - 390k_r^3 k_s^3 + 390k_r^4 k_s^2 - 120k_r^5 k_s\right) + \left(120k_r k_s^4 + 120k_r^2 k_s^3 - 630k_r^3 k_s^2 + 600k_r^4 k_s - 180k_r^5\right)\ \Big] + \ldots$$
$$\ldots + e^{-k_r t}\left(-150k_s^5 + 300k_r k_s^4 - 300k_r^2 k_s^3 + 150k_r^3 k_s^2 - 30k_r^4 k_s\right)\ \Big\}$$

**Equation 2**

Which in our hands was not amenable to further simplification.

*Analysis of petals intensity carpets:*

As discussed in the main text, the fluorescence intensity collected in the EGFP channel of a tracked active locus can be represented in the form of an intensity carpet, where each line represents the fluorescence intensity collected along the orbit (in most experiments 0 to 64 points corresponding to 0 to 360 degrees). Each line is collected at a time interval dictated by the temporal resolution of the experiment, and as illustrated in **Figure 1c**, the angular motion of each *petal* is reflected in a zigzag appearance of the intensity trace. In order to perform fluctuation analysis of the signal collected along the columns two corrections to the raw data need to be performed. First, the angular displacement of the trajectory over time needs to be corrected. We perform this by extracting sub-carpets that fully contain the intensity trace of only one petal. Each line of this petal carpet is fit by a 1D Gaussian Function to localize its center and the width of the trace. Initial guesses for the fit of line $k$ are estimated by calculating the fit over the integral of a set of $m$ preceding lines, from $k-m$ to $k$. This provides a memory or *inertia* to the fitting algorithm, as implemented in SimFCS (Globals Software, Enrico Gratton and the Laboratory for Fluorescence Dynamics). The typical memory $m$ that is used for the fit is 128 lines, corresponding to about 33 s at the typical acquisition frequency of 3.85 Hz.

Second, it is necessary to correct for trends in the fluorescence intensity over long time scales, such as photobleaching or decline in MS2-EGFP signal following transcription inhibition. However, as the signal intensity decays following perfusion of ActinomycinD (to inhibit transcription), it was necessary to correct the data. A moving average correction, calculated on the carpet of each petal and with a size of 1024 points=256 s was employed in this case.

*pCF analysis*

pCF was performed, as previously reported, using the SimFCS software, on the intensity carpet collected while performing tracking of a chromatin array. The voltage supplied to the galvanometer mirrors was adjusted in order to obtain a three-lobes shape, the trefoil illustrated in **Figure 2**, instead of a circular orbit.

*Phasor Analysis:*

Phasor analysis has been successfully used in the context of lifetime and spectral imaging, and we provide here for the first time an extension to the analysis of in-vivo enzyme kinetics.

Phasors of the experimental dataset were calculated from the petal intensity carpet according to the following strategy: each petal is divided into bins of 256 s (1024 points) and the fluorescence autocorrelation function (ACF) of each column of the petal-bin is calculated. The fluorescence ACFs calculated in each of the columns is then averaged to yield a global bin-petal ACF.

The phasor analysis is performed on this function as previously described [5]. Briefly, the FFT of the data is calculated and the real and imaginary parts of the first harmonic are used to calculate the modulus and the angle of a point in the complex space. In the current analysis, where we are ultimately interested in the rate constant, i.e. the slope of the autocorrelation function, only an angle is calculated, and therefore all the points lie on a unit-radius circle. Furthermore, this method acts as a high-pass filter on the data, and fluctuations on timescales longer than 256 s (<0.008 Hz) are effectively rejected. The process is schematically illustrated in **Supplementary Figure 6a-d**.

The phasor analysis of the autocorrelation curves as the one displayed in **Supplementary Figure 4c** is performed according to the standard equations introduced in the past for lifetime or spectral analysis [6], [5], [7]. For each autocorrelation curve described by the function *F(t)* a point on the unit circle in the complex plane is calculated using the *S* and *G* coordinates, according to

$$S(\omega_1) = \frac{\int F(t) \sin(\omega_1 t) dt}{\int F(t) dt}$$

$$G(\omega_1) = \frac{\int F(t) \cos(\omega_1 t) dt}{\int F(t) dt}$$

**Equation 3**

$\omega_1 = \frac{1}{2N\Delta t}$ is the fundamental frequency and is determined by the temporal resolution $\Delta t$ of our measurement and by the time bin used to calculate the ACF, typically 256 s yielding $N=1024$.

The angle $\phi$ of each *phasor* component is given by: $\phi = a \tan(\frac{S}{G})$. In practice $S$ and $G$ are calculated using a Fast Fourier Transform (FFT) of the Auto Correlation Function and posing:

$$G(k) = \text{Re} \frac{FFT(k) + FFT(N-k)}{(\sqrt{\text{Re}(FFT(k))^2 + \text{Im}(FFT(N-k))^2}}$$

$$S(k) = \text{Im} \frac{FFT(k) + FFT(N-k)}{(\sqrt{\text{Re}(FFT(k))^2 + \text{Im}(FFT(N-k))^2}}$$

**Equation 4**

where $k$ is the discrete frequency index corresponding to the continuous variable $\omega_1$ used in the previous formulas. $k=1$ yields the first harmonic of the measurement, $k=2$ the second harmonic and so forth.

In being a fit-less approach, phasor analysis requires a calibration to extract quantitative information from the data. We chose to calibrate the kinetic phasor using MonteCarlo generated PolII fluorescence trajectories.

It is worth noticing that the reference phasors depend upon the model and upon the parameters (such as initiation, elongation and termination constants) chosen to run the simulations. Upon changing the model, the absolute position of the experimental phasor would map to a different set of kinetic parameters; however, the observation of a relative difference between any two experimental datasets is not model dependent.

The obvious advantage of this approach is that it provides an immediate and graphical way to compare elongation rates across multiple datasets, capturing relative differences without the constraint of having to fit a model to the experimental data. We should also note here that elongation rates slower than about 10 bp/s and larger than about 200 bp/s are more difficult to resolve using the current binning interval of the fluorescence trajectories (1024 points=256 s) **Supplementary Figure 7c**, to the point that the 12 and 20 bp/s reference phasors almost overlap, as illustrated in **Supplementary Figure 7d**. Furthermore, the general trend of a clockwise increment of the elongation rates (reduction in phasor angle) holds true up to about 240 bp/s. This should be ascribed to the fact that above 240 bp/s the addition of new MS2 subunits goes beyond the temporal resolution of the measurement. Therefore in the current implementation the method appears to be able to resolve elongation rates over a one order of magnitude range between 20 bp/s and 240 bp/s.

By construction, all of the points fall on a circle (plus or minus a radial jitter to facilitate the display of the data), and the kinetic fingerprint of each of the petals is mapped into an angular value